\documentclass[reprint,prd,amsmath,amssymb,aps,floatfix,twocolumn]{revtex4-1}

\usepackage[english]{babel}
\usepackage{graphicx}
\usepackage{color}
\usepackage[dvipsnames]{xcolor}

\def\apj{ApJ}
\def\apjl{ApJ}

\def\jcap{JCAP}

\def\prd{Phys. Rev. D}
\def\mnras{MNRAS}

\begin{document}

\title{The inner engine of GeV-radiation-emitting gamma-ray bursts}

\author{R.~Ruffini,$^{1,2}$  J.~A.~Rueda,$^{1,2}$  R.~Moradi,$^{1,2}$ Y.~Wang,$^{1,2}$ S.~S.~Xue$^{1,2}$ L. Becerra,$^{1,3}$ C.~L.~Bianco,$^{1,2}$ Y.~C.~Chen,$^{1,2}$ C.~Cherubini,$^{4,5}$ S.~Filippi,$^{4,5}$ M.~Karlica,$^{1,2}$ J.~D.~Melon~Fuksman,$^{1,2}$  D.~Primorac,$^{1,2}$ N.~Sahakyan,$^{1,6}$ G.~V.~Vereshchagin,$^{1,2}$ }

\affiliation{$^1$International Center for Relativistic Astrophysics Network, Piazza della Repubblica 10, I-65122 Pescara, Italy\\
$^2$ICRA and Dipartimento di Fisica, Universit\`a  di Roma ``La Sapienza'', Piazzale Aldo Moro 5, I-00185 Roma, Italy\\
$^3$Escuela de F\'isica, Universidad Industrial de Santander, A.A.678, Bucaramanga, 680002, Colombia\\
$^4$Department of Engineering, University Campus Bio-Medico of Rome, Nonlinear Physics and Mathematical Modeling Lab, Via Alvaro del Portillo 21, 00128 Rome, Italy\\
$^5$International Center for Relativistic Astrophysics–ICRA, University Campus Bio-Medico of Rome, Via Alvaro del Portillo 21, I-00128 Rome, Italy\\
$^6$ICRANet-Armenia, Marshall Baghramian Avenue 24a, Yerevan 0019, Armenia.
}

\date{\today}

\begin{abstract}
We motivate how the most recent progress in the understanding the nature of the GeV radiation in most energetic gamma-ray bursts (GRBs), the binary-driven hypernovae (BdHNe), has led to the solution of a forty years unsolved problem in relativistic astrophysics: how to extract the rotational energy from a Kerr black hole for powering synchrotron emission and ultra high-energy cosmic rays. The \textit{inner engine} is identified in the proper use of a classical solution introduced by Wald in 1974 duly extended to the most extreme conditions found around the newborn black hole in a BdHN. The energy extraction process occurs in a sequence impulsive processes each accelerating protons to $10^{21}$~eV in a timescale of $10^{-6}$~s and in presence of an external magnetic field of $10^{14}$~G. Specific example is given for a black hole of initial angular momentum $J=0.3\,M^2$ and mass $M\approx 3\,M_\odot$ leading to the GeV radiation of $10^{49}$~erg$\cdot$s$^{-1}$. The process can energetically continue for thousands of years.
\end{abstract}


\maketitle

\section{Introduction}\label{sec:1}

\begin{figure}
    \centering
    \includegraphics[width=\hsize,clip]{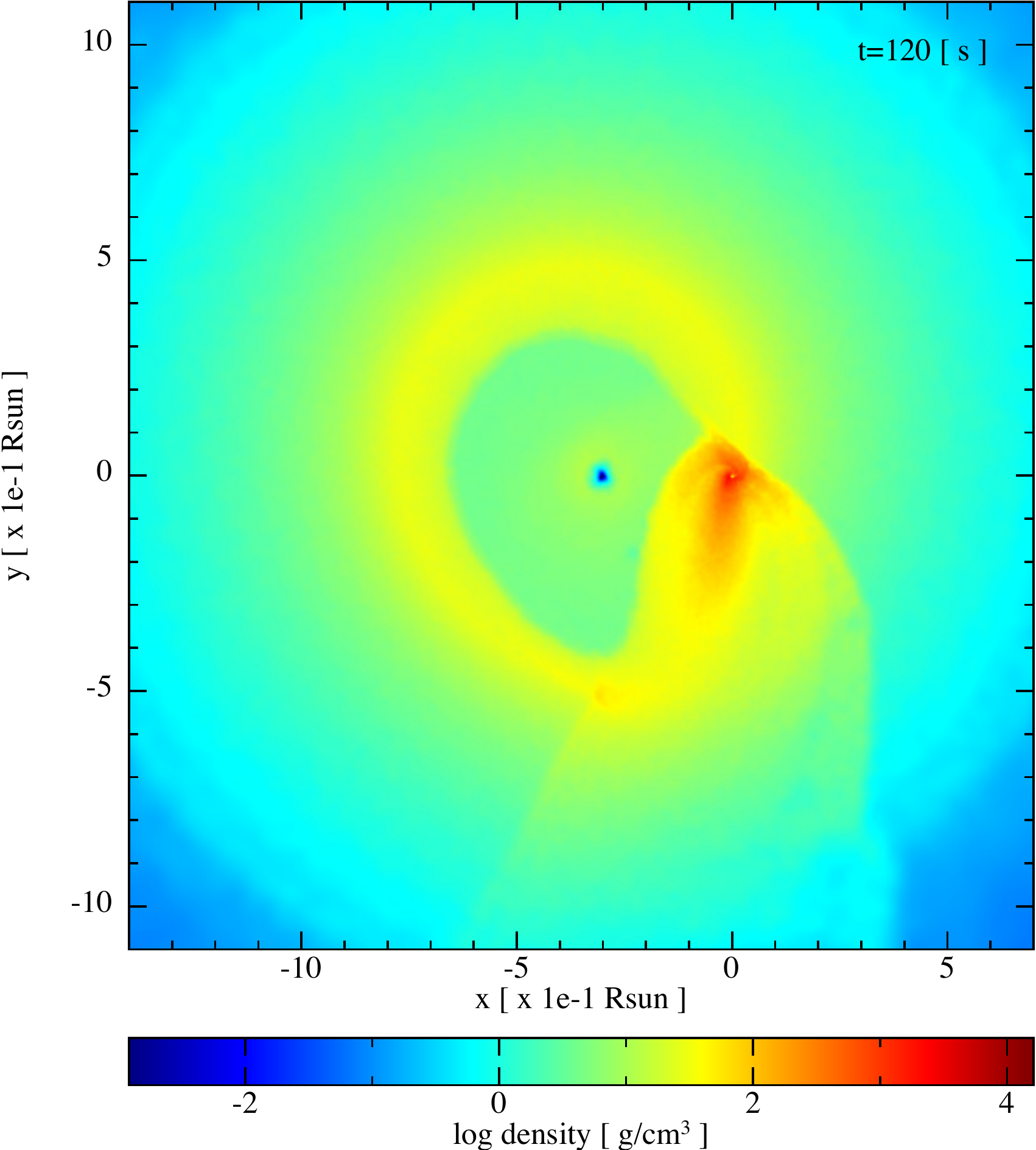}
    \caption{Selected SPH simulations from \cite{2019ApJ...871...14B} of the exploding CO$_{\rm core}$ as SN in presence of a companion NS: Model `25m1p08e' with $P_{\rm orb}=4.8$~min. The CO$_{\rm core}$ is taken from the $25~M_\odot$ zero-age main-sequence (ZAMS) progenitor, so it has a mass $M_{\rm CO}=6.85~M_\odot$. The mass of the NS companion is $M_{\rm NS}=2~M_\odot$. It is shown the density profile on the equatorial plane; the coordinate system has been rotated and translated in such a way that the NS companion is at the origin and the $\nu$NS is along the -x axis. The snapshot is at the time of the gravitational collapse of the NS companion to a BH, $t=120$~s from the SN shock breakout ($t=0$ of our simulation). The system forms a new binary system composed by the $\nu$NS (at the center of the deep-blue region) and the BH formed by the collapse of the NS companion (at the center of the red vortices). {The SN ejecta is the ionized medium that supplies the BH with the charged particles needed for the realization of the elementary processes leading to the Wald's solution.}}
    \label{fig:SPHsimulation}
\end{figure}

The understanding of gamma-ray bursts (GRBs), ongoing since almost fifty years, has been recently modified by the introduction of eight different subclasses of GRBs with binary progenitors composed of different combinations of neutron stars (NSs), carbon-oxygen cores (CO$_{\rm core}$) leading to supernovae (SNe), white dwarfs (WDs), and black holes (BHs), here indicated as Papers I.1 \cite{2016ApJ...832..136R,2018ApJ...859...30R,2018JCAP...10..006R1}. This approach contrast with  the traditional ones which assume all GRBs originating from a single BH with a {jetted blast-wave emission} extending from the ultrarelativistic prompt emission phase (UPE) all the way to the late observable GRB phases. Special attention has been given to identify the different components of a special class of GRBs originating from a tight binary system, of orbital period $\sim 5$~min, composed of a CO$_{\rm core}$, undergoing a SN event, and a NS companion. We have called these systems binary-driven hypernovae (BdHNe). In particular, in Papers I.2 \cite{2014ApJ...793L..36F,2015PhRvL.115w1102F,2015ApJ...812..100B,2016ApJ...833..107B,2018ApJ...852..120B} we have studied the physics of the hypercritical accretion of the SN ejecta onto the companion NS, and in Papers I.3 \cite{2016ApJ...833..107B,2019ApJ...871...14B} we have visualized by three-dimensional (3D) smoothed-particle-hydrodynamics (SPH) simulations the process of BH formation (see Fig.~\ref{fig:SPHsimulation}). From these works it has been clearly evidenced that the BH formation does not occur, as often idealized, in an asymptotically flat empty space but, it is surrounded by plasma undergoing quantum and classic ultrarelativistic regimes. {A cavity extending $\approx 10^{11}$~cm and of density $\sim 10^{-6}$~g~cm$^{-3}$ is formed around the BH \cite{2018ApJ...852..120B,2019ApJ...871...14B,2019arXiv190403163R}. The evolution of this cavity after the GRB explosion is addressed in \cite{2019arXiv190403163R} which shows that the density inside the cavity can be as low as $10^{-13}$~g~cm$^{-3}$.}

In 2018, we have further evidenced in Paper II.1 \cite{2018ApJ...869..151R} that in the GRB, itself a BdHN, the UPE, originating from the self-accelerating electron-positron ($e^-e^+$) plasma reaching Lorentz factor $\Gamma\sim 500$, is constituted by two spikes separated by $\sim 20$~s. Moreover, the impact of such a UPE on the SN ejecta gives origin to the gamma-ray flares and to the X-ray flares, and to an extended X-ray emission (EXE). Such an EXE occurs during the acceleration phase from the SN ejecta into a hypernova (HN). All these components have $\Gamma\lesssim 4$. In Paper II.2 \cite{2018ApJ...852...53R} we have given the basis for the model independent determination of the Lorentz factor $\Gamma\lesssim 2$ in the X-ray flares during the GRB plateau phase with explicit computation performed on $16$ BdHNe. In Paper II.3 \cite{2018ApJ...869..101R} we developed an afterglow description consistent with a mildly-relativistic expansion for the afterglow of GRB 130427A. There we have also evidenced the specific role of the pulsar-like behaviour of the $\nu$NS, originated from the SN explosion of the CO$_{\rm core}$ and fitted the observed synchrotron emission spectra of the afterglow within a mildly-relativistic expansion with $\Gamma\sim 2$.

From the above, it clearly follows that the traditional approach of assuming a single ultrarelativistic phase extending from the UPE to the late GeV-emission phase (GeV-P) is untenable. The strong limit imposed by the {intervening} mildly-relativistic expansion of the plateau and afterglow phase request {also} an alternative approach of the GeV-P to the currently accepted one.

{These factors have motivated our decision to address the origin of the GeV-P with a new energy source using the BH physics, following the UPE phase. It is indeed important to emphasize the fundamental difference between the UPE and GeV-P. The correlation between luminosity and variability in long GRBs is well known since the BATSE era (see e.g. \cite{1999ApJ...510..312S}) and it addresses the time variability in the MeV radiation in the UPE of the GRB. The duration of the UPE is of the order of of $T_{90} = 10$--$100$~s; the typical luminosity is $L_{\rm UPE}\sim E_{\rm iso}/T_{90}\sim 10^{53}/T_{90}\sim 10^{51}$--$10^{52}$~erg~s$^{-1}$ and the characteristic variability timescale is from a fraction of a second up to the order of $T_{90}$. Instead we are addressing a different process: the high-energy radiation in the GeV and TeV after the UPE. The emission process occurs in the ``quantized'' elementary events introduced in this article, namely $10^{44}$~erg emitted over a timescale of $10^{-6}$~s leading to a luminosity $10^{50}$~erg~s$^{-1}$. Moreover, this emission process is expected to last for thousands of years.}

We have shown in {Paper III.1 \cite{2018arXiv180305476R}} a first result: that the rotational energy of a Kerr BH is sufficient to explain the entire GeV-P. From this result, in turn, we have determine the mass and the spin of the BH using the data of the observed GeV luminosity of $21$ BdHNe. Using energy conservation argument, we have determined the slowing down rate of the Kerr BH. Specifically for GRB 130427A we have obtained
\begin{equation}\label{eq:BHparameters}
a=0.3\,M,\quad M=2.28 M_{\odot},
\end{equation}
which lead to a luminosity of $\sim 10^{50}$~erg~s$^{-1}$ and (see Tables 5 and 8 in Paper III.1 for details). Here $M$ and $a=J/M$ are the total mass and specific angular momentum of the Kerr BH, being $J$ its angular momentum.

For obtaining these results we have used the mass-energy formula of the Kerr BH, \citet{1970PhRvL..25.1596C}, and the one of the Kerr-Newman BH, \citet{1971PhRvD...4.3552C} and \citet{1971PhRvL..26.1344H, Hawking:1971vc}, 
\begin{subequations}
\renewcommand{\theequation}{\theparentequation.\arabic{equation}}
\begin{align}
\label{aone}
M^2 = \frac{J^2}{4 M^2_{irr}}+\left(\frac{Q^2}{4 M_{\rm irr}}+M_{\rm irr}\right)^2,\\
S = 16 \pi M^2_{\rm irr},\\ 
\delta S = 32 \pi  M_{\rm irr} \delta M_{\rm irr}\geq 0,
\end{align}
\end{subequations}
where $Q$ and $M_{\rm irr}$ and $S$ are the charge, irreducible mass and horizon surface area of the BH (in $c=G=1$ units).

It is then clear from the above that the BH rotational energy is able to power the GeV-P. It remains to answer
the crucial long-lasting 50 years old question of how to extract it through an efficient process that could explain the enormous luminosity of the most powerful astrophysical systems, AGN and GRBs.

We turn now here, first to the physical quantum and classical electrodynamics relativistic model able to explain the luminosity and spectra of the observed GeV-P and the creation of the $e^+e^-$ plasma of the UPE by vacuum polarization process, in terms of the rotational energy of the Kerr BH. We attempt as well to direct attention to explain the origin of ultra high-energy cosmic rays (UHECRs). 


{
We are here interested in 
utilizing the BH mass-energy formula and following rigorously its electrodynamical implications in a system composed of a Kerr metric background and a uniform magnetic field close to the BH horizon: the \emph{inner engine}. The rotating BH in the presence of the magnetic field $B_0$ induces an electromagnetic field described by the Wald solution \cite{1974PhRvD..10.1680W}. The ejecta surrounding the BH supply the necessary amount of ionized matter that is accelerated to ultrarelativistic energies at expenses of the BH rotational energy. These particles in presence of the magnetic field generate synchrotron radiation which explains the GeV-P (see, e.g., \cite{2018arXiv181200354R,2019arXiv190404162R,2019arXiv190403163R}, for the application of this model to specific GRB sources).
}

{The first attempt to extract energy from a Kerr BH, in absence of charge, by a plasma accreting onto it was presented in \cite{1975PhRvD..12.2959R}.  The infinite conductivity condition, $F_{\mu \nu}u^{\nu}= 0$ was there used leading to $E\cdot B= 0$. Under those conditions no process of energy extraction was there possible. The idea of using the theory of \textit{gaps}, following the seminal ideas in pulsar theory \cite{1971ApJ...164..529S,1975ApJ...196...51R}, to overcome the condition $E\cdot B= 0$ in the magnetosphere so the be able to accelerate particles in an accreting Kerr BH was introduced in \cite{1977MNRAS.179..433B}. It was there imposed a force-free condition, $F_{\rm \mu \nu}J^{\rm \nu}=0$, and introduced ``\textit{gaps}'' outside the BH horizon.} 

{As we mentioned above, our model for the high-energy emission is based on the solution by Wald solution. The electromagnetic field of this solution of the Einstein-Maxwell equations is characterized by $E\cdot B\neq 0$ \cite{1974PhRvD..10.1680W}. What we are here proposing is that this feature naturally allows the acceleration of particles, without the introduction of any \textit{gaps}. Consequently, it can be used as the basis of the GRB high-energy emission energetic requirements. In this work we focus on the beginning of the emission, the first emission event, and the extension of this work to the sequence, the evolution of the system by the repetition/intermittency of these events can be found in \cite{2018arXiv181200354R}.}

Therefore, our approach here goes directly to the heart of the profound difference of the electrodynamical process from any gravitational one, and an additional novelty is its application {of this alternative approach to the} GRB observations.

The article is organized as follows. In Sec.~\ref{sec:2} we summarize the structure of the electromagnetic field around the BH. Section~\ref{sec:3} is devoted to the analysis of the energy budget of the system for accelerating protons along the symmetry (rotation) axis, which becomes a source of UHECRs. We define the timescale of this elementary process and infer how long it can be at work by evaluating the rate at which it extracts the BH angular momentum. In Sec.~\ref{sec:4} we analyze the radiation losses by synchrotron emission that come from the off-axis proton motion and discuss the conditions under which it leads to GeV emission. Finally, Sec.~\ref{sec:5} we discuss the physical and astrophysical consequences of our results.

{\section{Electromagnetic field around the black hole}}\label{sec:2}

We now assume that the \textit{inner engine} of the GRB originates in the electrodynamical properties of a Papapetrou-Wald-Gibbons field \cite{1966AIHPA...4...83P,1974PhRvD..10.1680W,2013arXiv1301.3927G} which occurs when ``a stationary axisymmetric black hole which is placed in an originally uniform magnetic (test) field of strength $B_0$ aligned along the axis of symmetry of the black hole'' \cite{1974PhRvD..10.1680W}. The field $B_0$ is assumed to be constant in time.

We assume this \textit{inner engine} be surrounded by  1) an ionized plasma here, composed for simplicity, by protons and electrons, although the results can be easily generalized to the case of ions. 
2) The external sources of magnetic field $B_0$ originated from e.g.~by the pulsar-like {or multipole, e.g. toroidal, magnetic field of the $\nu$NS \cite{2018ApJ...869..101R,2019ApJ...874...39W}, or by the fossil field of the collapsed NS that led to the BH}. In either case, such sources should be located within a radius $R= 1/B_0$, in order to avoid unnecessary complications \cite{2013CQGra..30l5008G}. 3) The HN ejecta in the equatorial plane of the binary progenitor (see e.g.~Fig.~\ref{fig:SPHsimulation}, also Fig.~1 in paper II.1 and Refs.~\cite{2016ApJ...833..107B,2019ApJ...871...14B}). {This ionized medium supply the BH with the charged particles needed to establish the elementary process that lead to the Wald's solution.}

The Kerr spacetime metric (geometric units are considered), which is stationary and axisymmetric, in standard Boyer-Lindquist coordinates reads \cite{Misner}
\begin{eqnarray}
\label{Kmetric}
ds^2&=&-\left( 1- \frac{2Mr}{\Sigma} \right) dt^2-\frac{ 4aMr\sin^2\theta }{\Sigma} dt d\phi +\frac{\Sigma}{\Delta} dr^2\nonumber\\ &+&\Sigma d\theta^2+
\left[r^2+a^2+\frac{2Mra^2\sin^2\theta}{\Sigma}\right]\sin^2\theta
d\phi^2,
\end{eqnarray}
where $\Sigma=r^2+a^2\cos^2\theta$ and $\Delta=r^2-2Mr+a^2$. The  (outer) event horizon is located at $r_+=M+\sqrt{M^2-a^2}$.

The electromagnetic field of the \textit{inner engine} in orthonormal tetrad is:
\begin{align}
   E_{\hat{r}} &= \frac{a B_0}{\Sigma} \left(r\sin^2\theta-\frac{M \left(\cos ^2\theta+1\right) \left(r^2-a^2 \cos ^2\theta \right)}{\Sigma}\right),\\
    E_{\hat{\theta}}&=\frac{a B_0}{\Sigma}\sin\theta \cos\theta \sqrt{\Delta},\\ 
  B_{\hat{r}}&=-\frac{B_0 \cos\theta \left(-\frac{2 a^2 M r \left(\cos ^2\theta+1\right)}{\Sigma }+a^2+r^2\right)}{\Sigma },\\
  B_{\hat{\theta}}&=  \frac{B_0 r}{\Sigma}\sin\theta \sqrt{\Delta }.
 \end{align}

The most complex case of $\theta \neq 0$ will be considered elsewhere, we here for simplicity evaluate the field in the polar direction $\theta=0$:
\begin{eqnarray}
   E_{\hat{r}} &=& -B_0 \frac{2 J \left(r^2-a^2 \right)}{\left(r^2+a^2 \right)^2} \label{eq:ER} \\ 
    E_{\hat{\theta}}&=&0 \\
  B_{\hat{r}}&=&\frac{B_0  \left(-\frac{4 J a r}{\left(r^2+a^2 \right) }+a^2+r^2\right)}{\left(r^2+a^2 \right)}\\
  B_{\hat{\theta}}&=& 0.
 \end{eqnarray}
In order to evaluate the energetics of the \textit{inner engine} we introduce instead of Eq.~(\ref{eq:ER}) the new equation which, for the specific case of our GRBs, facilitates the comprehension of the nature of the \textit{inner engine}
\begin{eqnarray}\label{eq:ER2}
   E_{\hat{r}} &\approx& \frac{B_0 J}{2 M^2} \frac{r_+^2}{r^2}.
\end{eqnarray} 

In particular, Eq.~(\ref{eq:ER2}) allows to introduce only for the sake of a quantitative estimation, the \textit{effective charge}
\begin{eqnarray}\label{eq:EFCH}
  Q_{\rm eff}= \frac{ B_0 J}{2 M^2} r_+^2.
\end{eqnarray} 

\begin{figure}
\centering
\includegraphics[width=\hsize,clip]{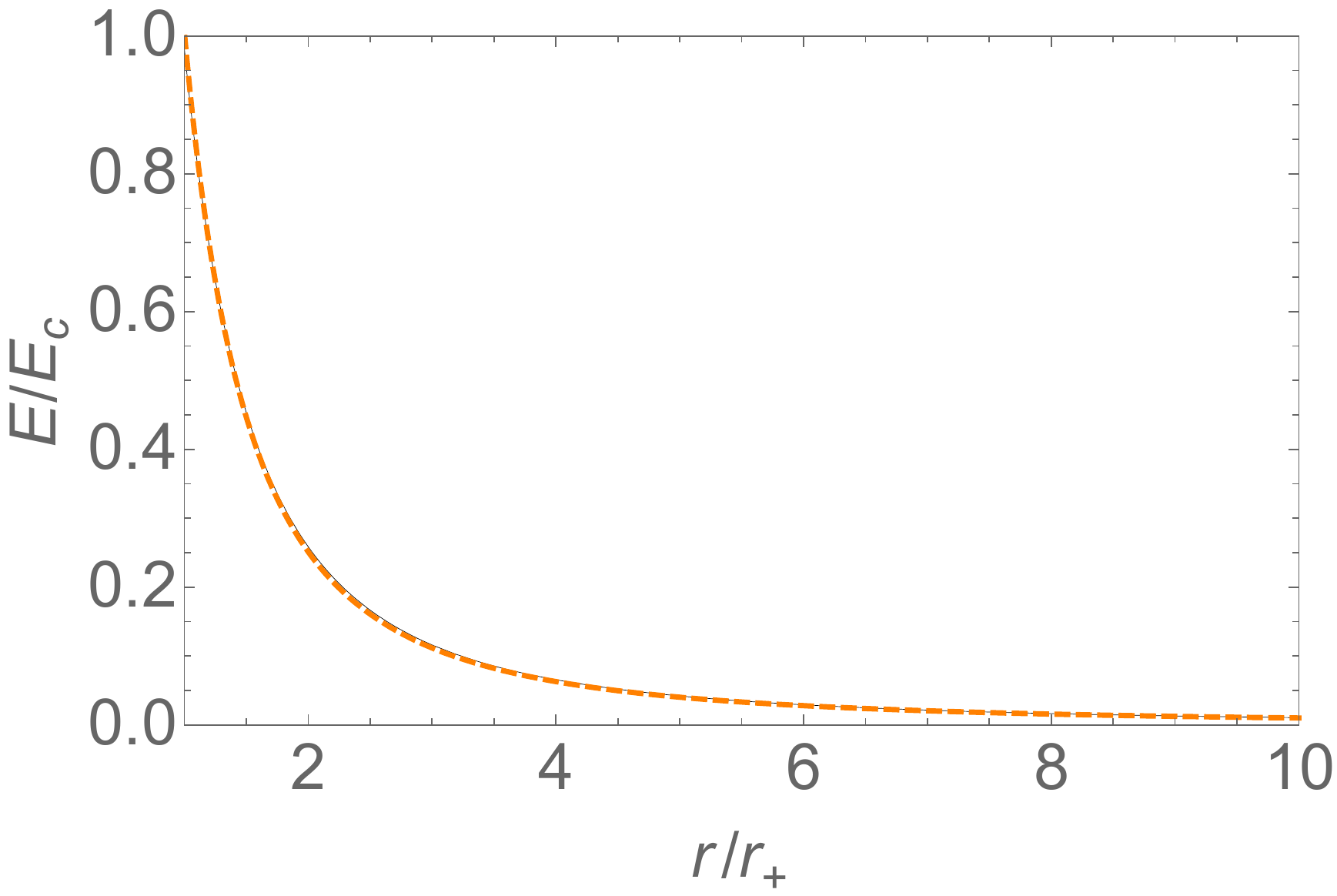}
\caption{Comparison of electric field given by the full expression (solid-black curve) given by Eq.~(\ref{eq:ER}) with the approximate expression (dashed-orange curve) given by Eq.~(\ref{eq:ER2}). Here we have used for the sake of example $a=0.3 M$ and $B_0/B_c \approx 6.7$, where $B_c = E_c$ being $E_c$ the critical field (\ref{eq:Ecrit}). The parameters $a$ and $B_0$ are such that the radial electric field at the horizon equals $E_c$. The quantitative relative difference between Eqs.~(\ref{eq:ER}) and (\ref{eq:ER2}) in this range of $r/r_+$ is only $2$--$4\%$.}
\label{fig:Ecomp}
\end{figure}

The two equations (\ref{eq:ER}) and (\ref{eq:ER2}) are in an excellent agreement, see Fig.~\ref{fig:Ecomp}, and lead to indistinguishable quantitative results in the case of 21 GRBs considered in Paper III.1. Equation (\ref{eq:ER2}) allows a more direct and straightforward evaluation of the leading terms. In the case $\theta\neq 0$ a multipolar distribution of charges should be considered. {Indeed, in \cite{2000NCimB.115..751M} the charge quadrupolar distribution in the Wald solution was explicitly shown and it was there indicated that acceleration of particles along the field lines would occur near the BH horizon.}

{\section{Elementary emission process and energy budget for UHECRs}}\label{sec:3}

Far from imposing any stationarity condition as often adopted in literature we examine a strongly time varying solution composed of a very large number of \textit{impulsive, elementary emissions}. Each \textit{impulsive emission} originates in presence of a magnetized ionized plasma from the discharge of an electric field $E$ with
 \begin{equation}\label{eq:Ecrit}
     E\lesssim E_c=\frac{m_e^2  c^3}{e\hbar},
 \end{equation}
where $E_{c}$ is the critical field, with  $m_e$ and $e$ the electron mass and charge, respectively.

We  assume that the magnetic field and the spin of the BH are antiparallel: protons in the surrounding ionized circumburst medium are repelled, while electrons are pulled in the BH.

Each impulsive event leads to a small loss of the rotational energy of the BH. All the attention of this Letter is directed to understand the physics of {the initial,} single elementary impulsive emission where protons are accelerated to the energy of the  $10^{21}$~eV. The sequence of \textit{impulsive elementary events} is reconstructed by an iterative process which takes into due account the loss of the rotational energy at the end of each impulse \cite{2018arXiv181200354R}.

The general, qualitative and quantitative features of the system can be obtained from a simple analysis and are as follows:

1) The rotating Kerr BH and the external uniform magnetic field $B_0$ induce an electric potential leading to a field near the horizon
\begin{align}\label{eq:Eh}
    E_{r_+}&\approx  \frac{J B_0}{2 M^2}=\frac{Q_{\rm eff}}{r_+^2} =\Omega_+ r_+ B_0 \nonumber\\
    &= 6.5\times 10^{15}\cdot \xi \beta\quad \frac{{\rm V}}{\rm cm},
\end{align}
where $\xi\equiv a/M = J/M^2$ and $\beta\equiv B_0/B_c$, $B_c=E_c$ where $E_c$ is the critical field (\ref{eq:Ecrit}) and $\Omega_+=a/(r_+^2+a^2)=a/(2 M r_+)$.

2) The associated electric potential difference, 
\begin{eqnarray}\label{eq:deltaphi}
    \Delta \phi &=& \frac{\epsilon_p}{e} = \int_{r_+}^\infty E dr = E_{r_+} r_+ \nonumber \\
    &=&9.7\times 10^{20}\cdot \xi \beta \mu (1+\sqrt{1-\xi^2})\quad\frac{{\rm V}}{e},
\end{eqnarray}
where $\mu\equiv M/M_\odot$, accelerates protons along the symmetry axis to Lorentz factors $\gamma_p = e \Delta\phi/(m_p c^2)\sim 10^{12}$, so to energies $\epsilon_p = \gamma_p m_p c^2 \sim 10^{21}$~eV.

3) For $\theta\neq 0$ the protons interact with the dense circumburst medium of the SN ejecta (see Fig.~\ref{fig:SPHsimulation}), reaching lower values of $\gamma_p$, e.g. via synchrotron process, leading to photons of energies $\epsilon_\gamma$ in the observed GeV range {(see \cite{2018arXiv181200354R}, for details)}.

4) The electrostatic energy available to accelerate protons is
\begin{equation}\label{eq:calE}
{\cal E} = \frac{1}{2} E_{r_+}^2 r_+^3 \approx 7.5\times 10^{41}\cdot\xi^2 \beta^2 \mu^3 (1+\sqrt{1-\xi^2})^3\quad{\rm erg},
\end{equation}
from which we can infer a number of protons
\begin{equation}\label{eq:Np}
    N_p = \frac{{\cal E}}{\epsilon_p} \approx 4.8\times 10^{32} \xi \beta \mu^2 (1+\sqrt{1-\xi^2})^2.
\end{equation}

5) The timescale of the elementary process is given by the time needed to accelerate the protons, i.e.
\begin{equation}\label{eq:deltaT}
    \Delta t_{\rm el} = \frac{\Delta \phi}{E_{r_+} c} = \frac{r_+}{c} \approx 4.9\times 10^{-6}\,\mu (1+\sqrt{1-\xi^2})\quad {\rm s}.
\end{equation}

6) Therefore, the electric power of the system is
\begin{equation}\label{eq:dotE}
    \dot{{\cal E}} \approx \frac{{\cal E}}{\Delta t_{\rm el}} = 1.5\times 10^{47}\cdot\xi^2 \beta^2 \mu^2 (1+\sqrt{1-\xi^2})^2\quad{\rm erg}\cdot {\rm s}^{-1},
\end{equation}

For example, in the case $\mu \approx 2.3$, $\xi = 0.3$, see Eq.~(\ref{eq:BHparameters}), and an electric field at the horizon equal to the critical field, which is obtained for $\beta = 2/\xi \approx 6.7$ (see Eq.~\ref{eq:Eh}), we obtain $N_p\approx 1.94\times 10^{34}$ protons and $\dot{{\cal E}} \approx 1.2\times 10^{49}$~erg~s$^{-1}$. This is in an excellent agreement with the GeV luminosity of the BdHN GRB 130427A after the prompt emission (see Paper III.1 for details). 

It is worth to recall that the above power has been estimated for the UHE protons escaping along the BH rotation axis which do not experience energy losses. That this electrostatic power be similar to the observed GeV luminosity gives great evidence that the protons which produce this emission have energies close to the highest energies of $10^{20}$~eV and they are accelerated off-axis (to experience synchrotron losses) but very close to it. This expectation is indeed confirmed by a detailed analysis of the synchrotron emission performed in \cite{2018arXiv181200354R} (see also next section).


In the elementary process timescale the BH experiences a fractional change of angular momentum
\begin{equation}
\frac{|\Delta J|}{J} = \frac{|\dot{J}|}{J}\Delta t_{\rm el} \sim \frac{|\Delta M|}{M} = \frac{|\dot{M}|}{M}\Delta t_{\rm el},
\end{equation}
which, using that the BH mass changes by $|\dot{M}| \approx L_{\rm GeV}$ (see Paper III.1), where $L_{\rm GeV}$ is the observed GeV luminosity, and using the same numbers we used above ($L_{\rm GeV} = 10^{49}$~erg~s$^{-1}$), leads to $|\Delta J|/J \approx 10^{-10}$. {Therefore, the single elementary emission event occurs with a reduction of $10^{-9}$ of the rotational energy of the Kerr BH (which is at most $29\%$ of $M$).} Then, the system starts over with a new elementary process, inducing a new electric field (\ref{eq:Eh}) now given by the new values of $J = J_0-\Delta J$ and $M = M_0-\Delta M$ and with $B$ unchanged since the field is external. The above estimates evidence the high efficiency of the present elementary process. In fact, if we assume that there are always enough particles surrounding the BH for the elementary process to be at work, the rotational energy budget of the BH can power it for thousands of years. It is clear that the actual duration of the sequence of elementary processes is limited by the amount of matter in the BH surroundings. We refer the reader to \cite{2018arXiv181200354R} for an extended discussion on this topic.

{\section{Off-axis radiation losses}}\label{sec:4}

The protons outside the symmetry axis experience energy losses, e.g. synchrotron emission. The proton Lorentz factor is limited by synchrotron losses to a value set by the balance between energy gain and energy loss per-unit-time, i.e. $\dot{E}_{\rm gain}=\dot{E}_{\rm loss}$, where {(see \cite{2018arXiv181200354R} for details):}
{
\begin{equation}\label{eq:Egain}
\dot{E}_{\rm gain} = e E_{r_+} c,\qquad \dot{E}_{\rm loss} = \frac{2}{3} \frac{e^4 B_0^2 \gamma^2\sin^2\left\langle\theta\right\rangle}{m_p^2 c^3},
\end{equation}
with $\left\langle\theta\right\rangle$ the average of the \emph{pitch angle}, i.e. the angle between the direction of the particle motion and the magnetic field.
}

The maximum proton Lorentz factor is thus:
{
\begin{equation}
\gamma_{\mathrm{max,p}}=\left(\frac{3\,\xi}{4\alpha\beta}\right)^{1/2}\frac{m_p/m_e}{\sin\left\langle \theta\right\rangle}\approx \frac{1.9\times 10^4}{\sin\left\langle \theta\right\rangle}\left(\frac{\xi}{\beta}\right)^{1/2},\label{eq:gmax}%
\end{equation}
where $\alpha$ is the fine structure constant. The spectrum of the proton-synchrotron emission peaks at the photon energy
\begin{eqnarray}
    \epsilon_{\rm max, \gamma} &=& \frac{e \hbar}{m_p c} B_0 \gamma_{\rm max,p}^2 = \frac{9}{8}\frac{m_p c^2}{\alpha}\frac{\xi}{\sin\left\langle \theta\right\rangle}\nonumber \\
    &\approx& 144\frac{\xi}{\sin\left\langle \theta\right\rangle}\,{\rm GeV},\label{eq:egamma}
\end{eqnarray}
and the luminosity is
\begin{equation}
    \dot{E}_{\rm loss} = \dot{E}_{\rm gain } = \frac{1}{2}e\,\xi B_0 \approx 3.2\times 10^{14}\,\xi\,\beta\quad {\rm erg}\cdot {\rm s}^{-1}.
\end{equation}
For the example numbers given by Eq.~(\ref{eq:BHparameters}) and $\beta\approx 6.7$, $\dot{E}_{\rm loss}\approx 6.4\times 10^{14}$~erg~s$^{-1}$.} From this we infer that, in order to obtain a luminosity of GeV photons of the order of few $10^{49}$~erg~s$^{-1}$, we would need about $10^{34}$ protons. 
This number of protons agrees with the available protons in the elementary process obtained from the energy budget. This is expected since using Eqs.~(\ref{eq:calE})--(\ref{eq:deltaT}) one finds that the available electrostatic power can be written as $\dot{\cal E} = N_p e E_{r_+} c$, which is the same as the maximum radiation loss per unit proton, $\dot{E}_{\rm loss}$ (see Eq.~\ref{eq:Egain}), multiplied by the total number of protons.

The analysis of the elementary process sequence is performed in a complementary work \cite{2018arXiv181200354R} and exemplified in the case of GRB 130427A. There, we first show from the observed GeV emission that the timescale of the elementary process increases with time following a precise linear law with time, i.e. the process becomes less efficient with time. Second, we show that this increasing timescale is explained by the evolution/decrease of the particle density of the HN ejecta around the BH site. The fit of the GeV emission observed and the expected UHECRs in the $21$ BdHNe in Paper III.1 is under computation and will be presented elsewhere.

{\section{Concluding remarks}}\label{sec:5}

We would like to stress that Eq.~(\ref{eq:ER2}) has been chosen only for mathematical and physical convenience to evaluate the leading orders with excellent approximation. However, the two equations remain conceptually very different. Equation Eq.~(\ref{eq:ER2}) can lead to an \textit{effective charge} interpretation of the phenomenon, but Eq.~(\ref{eq:ER}) corresponds to the real physical nature of the \textit{inner engine}. The two approaches quantitatively in an excellent approximation coincide but that small difference in Fig.~\ref{fig:Ecomp} is therefore crucial in differentiating the two conceptually very different physical processes. This can have paramount importance in the understanding of similarly fundamental physics issues.

It is interesting that the number density of surrounding ionized matter needed for the \emph{inner engine} to explain the GeV emission is much lower than the one requested for its explanation in terms of a traditional accretion disk process. The accretion process may produce a luminosity of the order of $\eta \dot{M} c^2$, where $\eta$ is the efficiency in converting the gravitational energy gained by accretion into radiation. Assuming a fiducial value of $\eta = 0.1$ (i.e. a $10\%$ of efficiency), a luminosity of $10^{49}$~erg~s$^{-1}$ requires an accretion rate of $\dot{M}\sim 10^{-4}~M_\odot$~s$^{-1}$. Thus, the BH must consume about $\dot{N}_p \sim \dot{M}/m_p\sim 10^{53}$~protons per second, or a proton number density rate $\dot{n}_p\sim 3 \dot{N}_p/(4\pi r_+^3)\sim 10^{35}$~cm$^{-3}$~s$^{-1}$. If the accretion efficiency $\eta$ is less than the fiducial $10\%$ adopted above, then the particle number required by accretion is still larger. Instead, our estimates imply that only $\dot{N}_p\sim N_p/\Delta t_{\rm el}\sim 10^{40}$~protons per second, or $\dot{n}_p\sim 10^{22}$~cm$^{-3}$~s$^{-1}$, are required by the present \emph{inner engine}. The above estimates explicitly show how the present electrodynamical process of BH rotational energy extraction is much more efficient than the gravitational process of accretion.

{Therefore, the \textit{electrodynamical accretion} here considered is characterized by densities much more tenuous with respect to the one characterizing the \textit{gravitational accretion}. The fully ionized plasma is here emitted in a much lower density cavity where the transparency conditions are much less severe. The low density cavity around the BH is crucial not only to fulfill the condition of low baryon density for the transparency and consequent observation of the MeV emission in the UPE, but also for the observation of the high-energy emission over the GeV domain discussed in this work (see \cite{2019arXiv190403163R,2019arXiv190404162R} for details).}

The considerations presented here are the preliminary concepts on how the \emph{inner engine} works. It becomes from now on a matter of interest to explore all the new physical and astrophysical implications derived from it, not only for GRBs but also for AGN. There are complementary complex phenomena which deserve to be pursued. For instance, the electromagnetic field of the Wald solution does not account for the feedback from the surrounding matter. The proton density required around the BH for the \emph{inner engine} to work is smaller than the Goldreich-Julian density \cite{1969ApJ...157..869G} by a factor $(3/2)(\hat{r}_+-\xi^2)^2/\hat{r}_+^3$, where $\hat{r}_+ = r_+/M$. 

In this line it is also worth to mention that the acceleration process and the consequent proton-synchrotron emission of high-energy photons occur on a short time and space scale so the elementary process presented here does not require that the magnetic field structure is kept over large scales far from the BH horizon.

{
There are most important aspects of the \emph{inner engine} presented here:
\begin{enumerate}
    \item 
    we have shown that the \emph{inner engine} powers a \emph{discrete elementary process} that accelerates particles which can either be source of UHECRs (see Sec.~\ref{sec:3}) or of synchrotron radiation (see Sec.~\ref{sec:4}) depending on the angle of their motion with respect to the magnetic field;
    \item
    the GeV-P is then explained by a series of such \emph{discrete, impulsive processes}, emitted on a timescale of $\sim 10^{-6}$~s and carrying an energy of $\sim 10^{44}$~erg;
    \item 
    {each of these single emission events in the present \textit{electrodynamical accretion} process occurs with a reduction of only $10^{-9}$ of the rotational energy of the Kerr BH, which implies a repetitive long-lasting nature of the emission, which is a unique feature very different from the traditional \textit{gravitational accretion} process (see e.g. \cite{2009MNRAS.394.2238V})};
    \item
    such a series of processes have recently found confirmation in the detailed time-resolved analysis of the light-curve and spectrum of the UPE and the GeV-P of GRB 190114C, which show self-similar and specific power-laws \cite{2019arXiv190404162R};
    \item
    the BH rotational energy reservoir can guarantee this sequence of \emph{impulsive} processes to continue for thousands of years, if the matter surrounding the BH provides the necessary amount of matter (see Sec.~\ref{sec:3} and \cite{2018arXiv181200354R}).
\end{enumerate} 
}

Before concluding, we would like to mention that the relevance of the Wald solution has been recently evidenced in the context of NS-BH mergers \cite{2018PhRvD..98l3002L}. It has been there shown that the dipole field of the NS at the BH horizon position can lead to an electromagnetic emission of interest in the approach to merger when the binary separation distance is very small and therefore the magnetic field at the BH site increases. 
Our considerations are markedly different: we do not involve a merger but the newly-formed $\nu$NS-BH binary from a BdHN, still embedded in the HN ejecta which is in our case the supplier of the charged particles needed to establish the elementary process and its associated timescale. In addition, the external magnetic field can reach higher values since it results from the fossil field of the NS which, by induced gravitational collapse, formed the BH \cite{2018ApJ...869..101R,2019ApJ...874...39W,2019arXiv190511339R}. As we have shown here these conditions of our system lead to a much energetic process which explains the high-energy emission of long GRBs. Indeed, the recent discovery of GRB 190114C and the determination of its redshift \citep{2019GCN.23695....1S} have allowed us to identify this new source as a BdHN I \citep{2019GCN.23715....1R}. The recent discovery by MAGIC of TeV radiation \citep{2019GCN.23701....1M} offers a testing ground for the theory presented in this article.



\begin{thebibliography}{39}
\expandafter\ifx\csname natexlab\endcsname\relax\def\natexlab#1{#1}\fi
\expandafter\ifx\csname bibnamefont\endcsname\relax
  \def\bibnamefont#1{#1}\fi
\expandafter\ifx\csname bibfnamefont\endcsname\relax
  \def\bibfnamefont#1{#1}\fi
\expandafter\ifx\csname citenamefont\endcsname\relax
  \def\citenamefont#1{#1}\fi
\expandafter\ifx\csname url\endcsname\relax
  \def\url#1{\texttt{#1}}\fi
\expandafter\ifx\csname urlprefix\endcsname\relax\def\urlprefix{URL }\fi
\providecommand{\bibinfo}[2]{#2}
\providecommand{\eprint}[2][]{\url{#2}}

\bibitem[{\citenamefont{{Ruffini} et~al.}(2016)\citenamefont{{Ruffini},
  {Rueda}, {Muccino}, {Aimuratov}, {Becerra}, {Bianco}, {Kovacevic}, {Moradi},
  {Oliveira}, {Pisani} et~al.}}]{2016ApJ...832..136R}
\bibinfo{author}{\bibfnamefont{R.}~\bibnamefont{{Ruffini}}},
  \bibinfo{author}{\bibfnamefont{J.~A.} \bibnamefont{{Rueda}}},
  \bibinfo{author}{\bibfnamefont{M.}~\bibnamefont{{Muccino}}},
  \bibinfo{author}{\bibfnamefont{Y.}~\bibnamefont{{Aimuratov}}},
  \bibinfo{author}{\bibfnamefont{L.~M.} \bibnamefont{{Becerra}}},
  \bibinfo{author}{\bibfnamefont{C.~L.} \bibnamefont{{Bianco}}},
  \bibinfo{author}{\bibfnamefont{M.}~\bibnamefont{{Kovacevic}}},
  \bibinfo{author}{\bibfnamefont{R.}~\bibnamefont{{Moradi}}},
  \bibinfo{author}{\bibfnamefont{F.~G.} \bibnamefont{{Oliveira}}},
  \bibinfo{author}{\bibfnamefont{G.~B.} \bibnamefont{{Pisani}}},
  \bibnamefont{et~al.}, \bibinfo{journal}{\apj} \textbf{\bibinfo{volume}{832}},
  \bibinfo{eid}{136} (\bibinfo{year}{2016}), \eprint{1602.02732}.

\bibitem[{\citenamefont{{Ruffini}
  et~al.}(2018{\natexlab{a}})\citenamefont{{Ruffini}, {Rodriguez}, {Muccino},
  {Rueda}, {Aimuratov}, {Barres de Almeida}, {Becerra}, {Bianco}, {Cherubini},
  {Filippi} et~al.}}]{2018ApJ...859...30R}
\bibinfo{author}{\bibfnamefont{R.}~\bibnamefont{{Ruffini}}},
  \bibinfo{author}{\bibfnamefont{J.}~\bibnamefont{{Rodriguez}}},
  \bibinfo{author}{\bibfnamefont{M.}~\bibnamefont{{Muccino}}},
  \bibinfo{author}{\bibfnamefont{J.~A.} \bibnamefont{{Rueda}}},
  \bibinfo{author}{\bibfnamefont{Y.}~\bibnamefont{{Aimuratov}}},
  \bibinfo{author}{\bibfnamefont{U.}~\bibnamefont{{Barres de Almeida}}},
  \bibinfo{author}{\bibfnamefont{L.}~\bibnamefont{{Becerra}}},
  \bibinfo{author}{\bibfnamefont{C.~L.} \bibnamefont{{Bianco}}},
  \bibinfo{author}{\bibfnamefont{C.}~\bibnamefont{{Cherubini}}},
  \bibinfo{author}{\bibfnamefont{S.}~\bibnamefont{{Filippi}}},
  \bibnamefont{et~al.}, \bibinfo{journal}{\apj} \textbf{\bibinfo{volume}{859}},
  \bibinfo{eid}{30} (\bibinfo{year}{2018}{\natexlab{a}}).

\bibitem[{\citenamefont{{Rueda} et~al.}(2018)\citenamefont{{Rueda}, {Ruffini},
  {Wang}, {Aimuratov}, {Barres de Almeida}, {Bianco}, {Chen}, {Lobato}, {Maia},
  {Primorac} et~al.}}]{2018JCAP...10..006R1}
\bibinfo{author}{\bibfnamefont{J.~A.} \bibnamefont{{Rueda}}},
  \bibinfo{author}{\bibfnamefont{R.}~\bibnamefont{{Ruffini}}},
  \bibinfo{author}{\bibfnamefont{Y.}~\bibnamefont{{Wang}}},
  \bibinfo{author}{\bibfnamefont{Y.}~\bibnamefont{{Aimuratov}}},
  \bibinfo{author}{\bibfnamefont{U.}~\bibnamefont{{Barres de Almeida}}},
  \bibinfo{author}{\bibfnamefont{C.~L.} \bibnamefont{{Bianco}}},
  \bibinfo{author}{\bibfnamefont{Y.~C.} \bibnamefont{{Chen}}},
  \bibinfo{author}{\bibfnamefont{R.~V.} \bibnamefont{{Lobato}}},
  \bibinfo{author}{\bibfnamefont{C.}~\bibnamefont{{Maia}}},
  \bibinfo{author}{\bibfnamefont{D.}~\bibnamefont{{Primorac}}},
  \bibnamefont{et~al.}, \bibinfo{journal}{\jcap} \textbf{\bibinfo{volume}{10}},
  \bibinfo{eid}{006} (\bibinfo{year}{2018}), \eprint{1802.10027}.

\bibitem[{\citenamefont{{Fryer} et~al.}(2014)\citenamefont{{Fryer}, {Rueda},
  and {Ruffini}}}]{2014ApJ...793L..36F}
\bibinfo{author}{\bibfnamefont{C.~L.} \bibnamefont{{Fryer}}},
  \bibinfo{author}{\bibfnamefont{J.~A.} \bibnamefont{{Rueda}}},
  \bibnamefont{and}
  \bibinfo{author}{\bibfnamefont{R.}~\bibnamefont{{Ruffini}}},
  \bibinfo{journal}{\apjl} \textbf{\bibinfo{volume}{793}}, \bibinfo{eid}{L36}
  (\bibinfo{year}{2014}), \eprint{1409.1473}.

\bibitem[{\citenamefont{{Fryer} et~al.}(2015)\citenamefont{{Fryer}, {Oliveira},
  {Rueda}, and {Ruffini}}}]{2015PhRvL.115w1102F}
\bibinfo{author}{\bibfnamefont{C.~L.} \bibnamefont{{Fryer}}},
  \bibinfo{author}{\bibfnamefont{F.~G.} \bibnamefont{{Oliveira}}},
  \bibinfo{author}{\bibfnamefont{J.~A.} \bibnamefont{{Rueda}}},
  \bibnamefont{and}
  \bibinfo{author}{\bibfnamefont{R.}~\bibnamefont{{Ruffini}}},
  \bibinfo{journal}{Physical Review Letters} \textbf{\bibinfo{volume}{115}},
  \bibinfo{eid}{231102} (\bibinfo{year}{2015}), \eprint{1505.02809}.

\bibitem[{\citenamefont{{Becerra} et~al.}(2015)\citenamefont{{Becerra},
  {Cipolletta}, {Fryer}, {Rueda}, and {Ruffini}}}]{2015ApJ...812..100B}
\bibinfo{author}{\bibfnamefont{L.}~\bibnamefont{{Becerra}}},
  \bibinfo{author}{\bibfnamefont{F.}~\bibnamefont{{Cipolletta}}},
  \bibinfo{author}{\bibfnamefont{C.~L.} \bibnamefont{{Fryer}}},
  \bibinfo{author}{\bibfnamefont{J.~A.} \bibnamefont{{Rueda}}},
  \bibnamefont{and}
  \bibinfo{author}{\bibfnamefont{R.}~\bibnamefont{{Ruffini}}},
  \bibinfo{journal}{\apj} \textbf{\bibinfo{volume}{812}}, \bibinfo{eid}{100}
  (\bibinfo{year}{2015}), \eprint{1505.07580}.

\bibitem[{\citenamefont{{Becerra} et~al.}(2016)\citenamefont{{Becerra},
  {Bianco}, {Fryer}, {Rueda}, and {Ruffini}}}]{2016ApJ...833..107B}
\bibinfo{author}{\bibfnamefont{L.}~\bibnamefont{{Becerra}}},
  \bibinfo{author}{\bibfnamefont{C.~L.} \bibnamefont{{Bianco}}},
  \bibinfo{author}{\bibfnamefont{C.~L.} \bibnamefont{{Fryer}}},
  \bibinfo{author}{\bibfnamefont{J.~A.} \bibnamefont{{Rueda}}},
  \bibnamefont{and}
  \bibinfo{author}{\bibfnamefont{R.}~\bibnamefont{{Ruffini}}},
  \bibinfo{journal}{\apj} \textbf{\bibinfo{volume}{833}}, \bibinfo{eid}{107}
  (\bibinfo{year}{2016}), \eprint{1606.02523}.

\bibitem[{\citenamefont{{Becerra} et~al.}(2018)\citenamefont{{Becerra},
  {Guzzo}, {Rossi-Torres}, {Rueda}, {Ruffini}, and
  {Uribe}}}]{2018ApJ...852..120B}
\bibinfo{author}{\bibfnamefont{L.}~\bibnamefont{{Becerra}}},
  \bibinfo{author}{\bibfnamefont{M.~M.} \bibnamefont{{Guzzo}}},
  \bibinfo{author}{\bibfnamefont{F.}~\bibnamefont{{Rossi-Torres}}},
  \bibinfo{author}{\bibfnamefont{J.~A.} \bibnamefont{{Rueda}}},
  \bibinfo{author}{\bibfnamefont{R.}~\bibnamefont{{Ruffini}}},
  \bibnamefont{and} \bibinfo{author}{\bibfnamefont{J.~D.}
  \bibnamefont{{Uribe}}}, \bibinfo{journal}{\apj}
  \textbf{\bibinfo{volume}{852}}, \bibinfo{eid}{120} (\bibinfo{year}{2018}),
  \eprint{1712.07210}.

\bibitem[{\citenamefont{{Becerra} et~al.}(2019)\citenamefont{{Becerra},
  {Ellinger}, {Fryer}, {Rueda}, and {Ruffini}}}]{2019ApJ...871...14B}
\bibinfo{author}{\bibfnamefont{L.}~\bibnamefont{{Becerra}}},
  \bibinfo{author}{\bibfnamefont{C.~L.} \bibnamefont{{Ellinger}}},
  \bibinfo{author}{\bibfnamefont{C.~L.} \bibnamefont{{Fryer}}},
  \bibinfo{author}{\bibfnamefont{J.~A.} \bibnamefont{{Rueda}}},
  \bibnamefont{and}
  \bibinfo{author}{\bibfnamefont{R.}~\bibnamefont{{Ruffini}}},
  \bibinfo{journal}{\apj} \textbf{\bibinfo{volume}{871}}, \bibinfo{eid}{14}
  (\bibinfo{year}{2019}), \eprint{1803.04356}.

\bibitem[{\citenamefont{{Ruffini}
  et~al.}(2019{\natexlab{a}})\citenamefont{{Ruffini}, {Melon Fuksman}, and
  {Vereshchagin}}}]{2019arXiv190403163R}
\bibinfo{author}{\bibfnamefont{R.}~\bibnamefont{{Ruffini}}},
  \bibinfo{author}{\bibfnamefont{J.~D.} \bibnamefont{{Melon Fuksman}}},
  \bibnamefont{and} \bibinfo{author}{\bibfnamefont{G.~V.}
  \bibnamefont{{Vereshchagin}}}, \bibinfo{journal}{arXiv e-prints}
  (\bibinfo{year}{2019}{\natexlab{a}}), \eprint{1904.03163}.

\bibitem[{\citenamefont{{Ruffini}
  et~al.}(2018{\natexlab{b}})\citenamefont{{Ruffini}, {Becerra}, {Bianco},
  {Chen}, {Karlica}, {Kova{\v c}evi{\'c}}, {Melon Fuksman}, {Moradi},
  {Muccino}, {Pisani} et~al.}}]{2018ApJ...869..151R}
\bibinfo{author}{\bibfnamefont{R.}~\bibnamefont{{Ruffini}}},
  \bibinfo{author}{\bibfnamefont{L.}~\bibnamefont{{Becerra}}},
  \bibinfo{author}{\bibfnamefont{C.~L.} \bibnamefont{{Bianco}}},
  \bibinfo{author}{\bibfnamefont{Y.~C.} \bibnamefont{{Chen}}},
  \bibinfo{author}{\bibfnamefont{M.}~\bibnamefont{{Karlica}}},
  \bibinfo{author}{\bibfnamefont{M.}~\bibnamefont{{Kova{\v c}evi{\'c}}}},
  \bibinfo{author}{\bibfnamefont{J.~D.} \bibnamefont{{Melon Fuksman}}},
  \bibinfo{author}{\bibfnamefont{R.}~\bibnamefont{{Moradi}}},
  \bibinfo{author}{\bibfnamefont{M.}~\bibnamefont{{Muccino}}},
  \bibinfo{author}{\bibfnamefont{G.~B.} \bibnamefont{{Pisani}}},
  \bibnamefont{et~al.}, \bibinfo{journal}{\apj} \textbf{\bibinfo{volume}{869}},
  \bibinfo{eid}{151} (\bibinfo{year}{2018}{\natexlab{b}}), \eprint{1712.05001}.

\bibitem[{\citenamefont{{Ruffini}
  et~al.}(2018{\natexlab{c}})\citenamefont{{Ruffini}, {Wang}, {Aimuratov},
  {Barres de Almeida}, {Becerra}, {Bianco}, {Chen}, {Karlica}, {Kovacevic},
  {Li} et~al.}}]{2018ApJ...852...53R}
\bibinfo{author}{\bibfnamefont{R.}~\bibnamefont{{Ruffini}}},
  \bibinfo{author}{\bibfnamefont{Y.}~\bibnamefont{{Wang}}},
  \bibinfo{author}{\bibfnamefont{Y.}~\bibnamefont{{Aimuratov}}},
  \bibinfo{author}{\bibfnamefont{U.}~\bibnamefont{{Barres de Almeida}}},
  \bibinfo{author}{\bibfnamefont{L.}~\bibnamefont{{Becerra}}},
  \bibinfo{author}{\bibfnamefont{C.~L.} \bibnamefont{{Bianco}}},
  \bibinfo{author}{\bibfnamefont{Y.~C.} \bibnamefont{{Chen}}},
  \bibinfo{author}{\bibfnamefont{M.}~\bibnamefont{{Karlica}}},
  \bibinfo{author}{\bibfnamefont{M.}~\bibnamefont{{Kovacevic}}},
  \bibinfo{author}{\bibfnamefont{L.}~\bibnamefont{{Li}}}, \bibnamefont{et~al.},
  \bibinfo{journal}{\apj} \textbf{\bibinfo{volume}{852}}, \bibinfo{eid}{53}
  (\bibinfo{year}{2018}{\natexlab{c}}), \eprint{1704.03821}.

\bibitem[{\citenamefont{{Ruffini}
  et~al.}(2018{\natexlab{d}})\citenamefont{{Ruffini}, {Karlica}, {Sahakyan},
  {Rueda}, {Wang}, {Mathews}, {Bianco}, and {Muccino}}}]{2018ApJ...869..101R}
\bibinfo{author}{\bibfnamefont{R.}~\bibnamefont{{Ruffini}}},
  \bibinfo{author}{\bibfnamefont{M.}~\bibnamefont{{Karlica}}},
  \bibinfo{author}{\bibfnamefont{N.}~\bibnamefont{{Sahakyan}}},
  \bibinfo{author}{\bibfnamefont{J.~A.} \bibnamefont{{Rueda}}},
  \bibinfo{author}{\bibfnamefont{Y.}~\bibnamefont{{Wang}}},
  \bibinfo{author}{\bibfnamefont{G.~J.} \bibnamefont{{Mathews}}},
  \bibinfo{author}{\bibfnamefont{C.~L.} \bibnamefont{{Bianco}}},
  \bibnamefont{and}
  \bibinfo{author}{\bibfnamefont{M.}~\bibnamefont{{Muccino}}},
  \bibinfo{journal}{\apj} \textbf{\bibinfo{volume}{869}}, \bibinfo{eid}{101}
  (\bibinfo{year}{2018}{\natexlab{d}}), \eprint{1712.05000}.

\bibitem[{\citenamefont{{Stern} et~al.}(1999)\citenamefont{{Stern}, {Poutanen},
  and {Svensson}}}]{1999ApJ...510..312S}
\bibinfo{author}{\bibfnamefont{B.}~\bibnamefont{{Stern}}},
  \bibinfo{author}{\bibfnamefont{J.}~\bibnamefont{{Poutanen}}},
  \bibnamefont{and}
  \bibinfo{author}{\bibfnamefont{R.}~\bibnamefont{{Svensson}}},
  \bibinfo{journal}{\apj} \textbf{\bibinfo{volume}{510}}, \bibinfo{pages}{312}
  (\bibinfo{year}{1999}), \eprint{astro-ph/9709113}.

\bibitem[{\citenamefont{{Ruffini}
  et~al.}(2018{\natexlab{e}})\citenamefont{{Ruffini}, {Moradi}, {Rueda},
  {Wang}, {Aimuratov}, {Becerra}, {Bianco}, {Chen}, {Cherubini}, {Filippi}
  et~al.}}]{2018arXiv180305476R}
\bibinfo{author}{\bibfnamefont{R.}~\bibnamefont{{Ruffini}}},
  \bibinfo{author}{\bibfnamefont{R.}~\bibnamefont{{Moradi}}},
  \bibinfo{author}{\bibfnamefont{J.~A.} \bibnamefont{{Rueda}}},
  \bibinfo{author}{\bibfnamefont{Y.}~\bibnamefont{{Wang}}},
  \bibinfo{author}{\bibfnamefont{Y.}~\bibnamefont{{Aimuratov}}},
  \bibinfo{author}{\bibfnamefont{L.}~\bibnamefont{{Becerra}}},
  \bibinfo{author}{\bibfnamefont{C.~L.} \bibnamefont{{Bianco}}},
  \bibinfo{author}{\bibfnamefont{Y.-C.} \bibnamefont{{Chen}}},
  \bibinfo{author}{\bibfnamefont{C.}~\bibnamefont{{Cherubini}}},
  \bibinfo{author}{\bibfnamefont{S.}~\bibnamefont{{Filippi}}},
  \bibnamefont{et~al.}, \bibinfo{journal}{ArXiv e-prints}
  (\bibinfo{year}{2018}{\natexlab{e}}), \eprint{1803.05476}.

\bibitem[{\citenamefont{{Christodoulou}}(1970)}]{1970PhRvL..25.1596C}
\bibinfo{author}{\bibfnamefont{D.}~\bibnamefont{{Christodoulou}}},
  \bibinfo{journal}{Physical Review Letters} \textbf{\bibinfo{volume}{25}},
  \bibinfo{pages}{1596} (\bibinfo{year}{1970}).

\bibitem[{\citenamefont{{Christodoulou} and
  {Ruffini}}(1971)}]{1971PhRvD...4.3552C}
\bibinfo{author}{\bibfnamefont{D.}~\bibnamefont{{Christodoulou}}}
  \bibnamefont{and}
  \bibinfo{author}{\bibfnamefont{R.}~\bibnamefont{{Ruffini}}},
  \bibinfo{journal}{\prd} \textbf{\bibinfo{volume}{4}}, \bibinfo{pages}{3552}
  (\bibinfo{year}{1971}).

\bibitem[{\citenamefont{{Hawking}}(1971)}]{1971PhRvL..26.1344H}
\bibinfo{author}{\bibfnamefont{S.~W.} \bibnamefont{{Hawking}}},
  \bibinfo{journal}{Physical Review Letters} \textbf{\bibinfo{volume}{26}},
  \bibinfo{pages}{1344} (\bibinfo{year}{1971}).

\bibitem[{\citenamefont{Hawking}(1972)}]{Hawking:1971vc}
\bibinfo{author}{\bibfnamefont{S.~W.} \bibnamefont{Hawking}},
  \bibinfo{journal}{Commun. Math. Phys.} \textbf{\bibinfo{volume}{25}},
  \bibinfo{pages}{152} (\bibinfo{year}{1972}).

\bibitem[{\citenamefont{{Wald}}(1974)}]{1974PhRvD..10.1680W}
\bibinfo{author}{\bibfnamefont{R.~M.} \bibnamefont{{Wald}}},
  \bibinfo{journal}{\prd} \textbf{\bibinfo{volume}{10}}, \bibinfo{pages}{1680}
  (\bibinfo{year}{1974}).

\bibitem[{\citenamefont{{Ruffini}
  et~al.}(2018{\natexlab{f}})\citenamefont{{Ruffini}, {Moradi}, {Rueda},
  {Becerra}, {Bianco}, {Cherubini}, {Filippi}, {Chen}, {Karlica}, {Sahakyan}
  et~al.}}]{2018arXiv181200354R}
\bibinfo{author}{\bibfnamefont{R.}~\bibnamefont{{Ruffini}}},
  \bibinfo{author}{\bibfnamefont{R.}~\bibnamefont{{Moradi}}},
  \bibinfo{author}{\bibfnamefont{J.~A.} \bibnamefont{{Rueda}}},
  \bibinfo{author}{\bibfnamefont{L.}~\bibnamefont{{Becerra}}},
  \bibinfo{author}{\bibfnamefont{C.~L.} \bibnamefont{{Bianco}}},
  \bibinfo{author}{\bibfnamefont{C.}~\bibnamefont{{Cherubini}}},
  \bibinfo{author}{\bibfnamefont{S.}~\bibnamefont{{Filippi}}},
  \bibinfo{author}{\bibfnamefont{Y.~C.} \bibnamefont{{Chen}}},
  \bibinfo{author}{\bibfnamefont{M.}~\bibnamefont{{Karlica}}},
  \bibinfo{author}{\bibfnamefont{N.}~\bibnamefont{{Sahakyan}}},
  \bibnamefont{et~al.}, \bibinfo{journal}{arXiv e-prints}
  (\bibinfo{year}{2018}{\natexlab{f}}), \eprint{1812.00354}.

\bibitem[{\citenamefont{{Ruffini}
  et~al.}(2019{\natexlab{b}})\citenamefont{{Ruffini}, {Li}, {Moradi}, {Rueda},
  {Wang}, {Xue}, {Bianco}, {Campion}, {Melon Fuksman}, {Cherubini}
  et~al.}}]{2019arXiv190404162R}
\bibinfo{author}{\bibfnamefont{R.}~\bibnamefont{{Ruffini}}},
  \bibinfo{author}{\bibfnamefont{L.}~\bibnamefont{{Li}}},
  \bibinfo{author}{\bibfnamefont{R.}~\bibnamefont{{Moradi}}},
  \bibinfo{author}{\bibfnamefont{J.~A.} \bibnamefont{{Rueda}}},
  \bibinfo{author}{\bibfnamefont{Y.}~\bibnamefont{{Wang}}},
  \bibinfo{author}{\bibfnamefont{S.~S.} \bibnamefont{{Xue}}},
  \bibinfo{author}{\bibfnamefont{C.~L.} \bibnamefont{{Bianco}}},
  \bibinfo{author}{\bibfnamefont{S.}~\bibnamefont{{Campion}}},
  \bibinfo{author}{\bibfnamefont{J.~D.} \bibnamefont{{Melon Fuksman}}},
  \bibinfo{author}{\bibfnamefont{C.}~\bibnamefont{{Cherubini}}},
  \bibnamefont{et~al.}, \bibinfo{journal}{arXiv e-prints}
  (\bibinfo{year}{2019}{\natexlab{b}}), \eprint{1904.04162}.

\bibitem[{\citenamefont{{Ruffini} and {Wilson}}(1975)}]{1975PhRvD..12.2959R}
\bibinfo{author}{\bibfnamefont{R.}~\bibnamefont{{Ruffini}}} \bibnamefont{and}
  \bibinfo{author}{\bibfnamefont{J.~R.} \bibnamefont{{Wilson}}},
  \bibinfo{journal}{\prd} \textbf{\bibinfo{volume}{12}}, \bibinfo{pages}{2959}
  (\bibinfo{year}{1975}).

\bibitem[{\citenamefont{{Sturrock}}(1971)}]{1971ApJ...164..529S}
\bibinfo{author}{\bibfnamefont{P.~A.} \bibnamefont{{Sturrock}}},
  \bibinfo{journal}{\apj} \textbf{\bibinfo{volume}{164}}, \bibinfo{pages}{529}
  (\bibinfo{year}{1971}).

\bibitem[{\citenamefont{{Ruderman} and
  {Sutherland}}(1975)}]{1975ApJ...196...51R}
\bibinfo{author}{\bibfnamefont{M.~A.} \bibnamefont{{Ruderman}}}
  \bibnamefont{and} \bibinfo{author}{\bibfnamefont{P.~G.}
  \bibnamefont{{Sutherland}}}, \bibinfo{journal}{\apj}
  \textbf{\bibinfo{volume}{196}}, \bibinfo{pages}{51} (\bibinfo{year}{1975}).

\bibitem[{\citenamefont{{Blandford} and {Znajek}}(1977)}]{1977MNRAS.179..433B}
\bibinfo{author}{\bibfnamefont{R.~D.} \bibnamefont{{Blandford}}}
  \bibnamefont{and} \bibinfo{author}{\bibfnamefont{R.~L.}
  \bibnamefont{{Znajek}}}, \bibinfo{journal}{\mnras}
  \textbf{\bibinfo{volume}{179}}, \bibinfo{pages}{433} (\bibinfo{year}{1977}).

\bibitem[{\citenamefont{{Papapetrou}}(1966)}]{1966AIHPA...4...83P}
\bibinfo{author}{\bibfnamefont{A.}~\bibnamefont{{Papapetrou}}},
  \bibinfo{journal}{Annales de L'Institut Henri Poincare Section (A) Physique
  Theorique} \textbf{\bibinfo{volume}{4}}, \bibinfo{pages}{83}
  (\bibinfo{year}{1966}).

\bibitem[{\citenamefont{{Gibbons}
  et~al.}(2013{\natexlab{a}})\citenamefont{{Gibbons}, {Mujtaba}, and
  {Pope}}}]{2013arXiv1301.3927G}
\bibinfo{author}{\bibfnamefont{G.~W.} \bibnamefont{{Gibbons}}},
  \bibinfo{author}{\bibfnamefont{A.~H.} \bibnamefont{{Mujtaba}}},
  \bibnamefont{and} \bibinfo{author}{\bibfnamefont{C.~N.}
  \bibnamefont{{Pope}}}, \bibinfo{journal}{ArXiv e-prints}
  (\bibinfo{year}{2013}{\natexlab{a}}), \eprint{1301.3927}.

\bibitem[{\citenamefont{{Wang} et~al.}(2019)\citenamefont{{Wang}, {Rueda},
  {Ruffini}, {Becerra}, {Bianco}, {Becerra}, {Li}, and
  {Karlica}}}]{2019ApJ...874...39W}
\bibinfo{author}{\bibfnamefont{Y.}~\bibnamefont{{Wang}}},
  \bibinfo{author}{\bibfnamefont{J.~A.} \bibnamefont{{Rueda}}},
  \bibinfo{author}{\bibfnamefont{R.}~\bibnamefont{{Ruffini}}},
  \bibinfo{author}{\bibfnamefont{L.}~\bibnamefont{{Becerra}}},
  \bibinfo{author}{\bibfnamefont{C.}~\bibnamefont{{Bianco}}},
  \bibinfo{author}{\bibfnamefont{L.}~\bibnamefont{{Becerra}}},
  \bibinfo{author}{\bibfnamefont{L.}~\bibnamefont{{Li}}}, \bibnamefont{and}
  \bibinfo{author}{\bibfnamefont{M.}~\bibnamefont{{Karlica}}},
  \bibinfo{journal}{\apj} \textbf{\bibinfo{volume}{874}}, \bibinfo{eid}{39}
  (\bibinfo{year}{2019}), \eprint{1811.05433}.

\bibitem[{\citenamefont{{Gibbons}
  et~al.}(2013{\natexlab{b}})\citenamefont{{Gibbons}, {Mujtaba}, and
  {Pope}}}]{2013CQGra..30l5008G}
\bibinfo{author}{\bibfnamefont{G.~W.} \bibnamefont{{Gibbons}}},
  \bibinfo{author}{\bibfnamefont{A.~H.} \bibnamefont{{Mujtaba}}},
  \bibnamefont{and} \bibinfo{author}{\bibfnamefont{C.~N.}
  \bibnamefont{{Pope}}}, \bibinfo{journal}{Classical and Quantum Gravity}
  \textbf{\bibinfo{volume}{30}}, \bibinfo{eid}{125008}
  (\bibinfo{year}{2013}{\natexlab{b}}).

\bibitem[{\citenamefont{Misner et~al.}(1973)\citenamefont{Misner, Thorne, and
  Wheeler.}}]{Misner}
\bibinfo{author}{\bibfnamefont{C.~W.} \bibnamefont{Misner}},
  \bibinfo{author}{\bibfnamefont{K.~S.} \bibnamefont{Thorne}},
  \bibnamefont{and} \bibinfo{author}{\bibfnamefont{J.~A.}
  \bibnamefont{Wheeler.}}, \emph{\bibinfo{title}{{Gravitation}}}
  (\bibinfo{year}{1973}).

\bibitem[{\citenamefont{{Miniutti} and {Ruffini}}(2000)}]{2000NCimB.115..751M}
\bibinfo{author}{\bibfnamefont{G.}~\bibnamefont{{Miniutti}}} \bibnamefont{and}
  \bibinfo{author}{\bibfnamefont{R.}~\bibnamefont{{Ruffini}}},
  \bibinfo{journal}{Nuovo Cimento B Serie} \textbf{\bibinfo{volume}{115}},
  \bibinfo{pages}{751} (\bibinfo{year}{2000}).

\bibitem[{\citenamefont{{Goldreich} and {Julian}}(1969)}]{1969ApJ...157..869G}
\bibinfo{author}{\bibfnamefont{P.}~\bibnamefont{{Goldreich}}} \bibnamefont{and}
  \bibinfo{author}{\bibfnamefont{W.~H.} \bibnamefont{{Julian}}},
  \bibinfo{journal}{\apj} \textbf{\bibinfo{volume}{157}}, \bibinfo{pages}{869}
  (\bibinfo{year}{1969}).

\bibitem[{\citenamefont{{van Putten} and {Gupta}}(2009)}]{2009MNRAS.394.2238V}
\bibinfo{author}{\bibfnamefont{M.~H.~P.~M.} \bibnamefont{{van Putten}}}
  \bibnamefont{and} \bibinfo{author}{\bibfnamefont{A.~C.}
  \bibnamefont{{Gupta}}}, \bibinfo{journal}{\mnras}
  \textbf{\bibinfo{volume}{394}}, \bibinfo{pages}{2238} (\bibinfo{year}{2009}),
  \eprint{0901.1674}.

\bibitem[{\citenamefont{{Levin} et~al.}(2018)\citenamefont{{Levin}, {D'Orazio},
  and {Garcia-Saenz}}}]{2018PhRvD..98l3002L}
\bibinfo{author}{\bibfnamefont{J.}~\bibnamefont{{Levin}}},
  \bibinfo{author}{\bibfnamefont{D.~J.} \bibnamefont{{D'Orazio}}},
  \bibnamefont{and}
  \bibinfo{author}{\bibfnamefont{S.}~\bibnamefont{{Garcia-Saenz}}},
  \bibinfo{journal}{\prd} \textbf{\bibinfo{volume}{98}}, \bibinfo{eid}{123002}
  (\bibinfo{year}{2018}), \eprint{1808.07887}.

\bibitem[{\citenamefont{{Rueda} et~al.}(2019)\citenamefont{{Rueda}, {Ruffini},
  {Karlica}, {Moradi}, and {Wang}}}]{2019arXiv190511339R}
\bibinfo{author}{\bibfnamefont{J.~A.} \bibnamefont{{Rueda}}},
  \bibinfo{author}{\bibfnamefont{R.}~\bibnamefont{{Ruffini}}},
  \bibinfo{author}{\bibfnamefont{M.}~\bibnamefont{{Karlica}}},
  \bibinfo{author}{\bibfnamefont{R.}~\bibnamefont{{Moradi}}}, \bibnamefont{and}
  \bibinfo{author}{\bibfnamefont{Y.}~\bibnamefont{{Wang}}},
  \bibinfo{journal}{arXiv e-prints}  (\bibinfo{year}{2019}),
  \eprint{1905.11339}.

\bibitem[{\citenamefont{{Selsing} et~al.}(2019)\citenamefont{{Selsing},
  {Fynbo}, {Heintz}, and {Watson}}}]{2019GCN.23695....1S}
\bibinfo{author}{\bibfnamefont{J.}~\bibnamefont{{Selsing}}},
  \bibinfo{author}{\bibfnamefont{J.~P.~U.} \bibnamefont{{Fynbo}}},
  \bibinfo{author}{\bibfnamefont{K.~E.} \bibnamefont{{Heintz}}},
  \bibnamefont{and} \bibinfo{author}{\bibfnamefont{D.}~\bibnamefont{{Watson}}},
  \bibinfo{journal}{GRB Coordinates Network, Circular Service, No.~23695, \#1
  (2019)} \textbf{\bibinfo{volume}{23695}} (\bibinfo{year}{2019}).

\bibitem[{\citenamefont{{Ruffini}
  et~al.}(2019{\natexlab{c}})\citenamefont{{Ruffini}, {Moradi}, {Aimuratov},
  {Barres}, {Belinski}, {Bianco}, {Chen}, {Cherubini}, {Filippi}, {Fuksman}
  et~al.}}]{2019GCN.23715....1R}
\bibinfo{author}{\bibfnamefont{R.}~\bibnamefont{{Ruffini}}},
  \bibinfo{author}{\bibfnamefont{R.}~\bibnamefont{{Moradi}}},
  \bibinfo{author}{\bibfnamefont{Y.}~\bibnamefont{{Aimuratov}}},
  \bibinfo{author}{\bibfnamefont{U.}~\bibnamefont{{Barres}}},
  \bibinfo{author}{\bibfnamefont{V.~A.} \bibnamefont{{Belinski}}},
  \bibinfo{author}{\bibfnamefont{C.~L.} \bibnamefont{{Bianco}}},
  \bibinfo{author}{\bibfnamefont{Y.~C.} \bibnamefont{{Chen}}},
  \bibinfo{author}{\bibfnamefont{C.}~\bibnamefont{{Cherubini}}},
  \bibinfo{author}{\bibfnamefont{S.}~\bibnamefont{{Filippi}}},
  \bibinfo{author}{\bibfnamefont{D.~M.} \bibnamefont{{Fuksman}}},
  \bibnamefont{et~al.}, \bibinfo{journal}{GRB Coordinates Network, Circular
  Service, No.~23715, \#1 (2019)} \textbf{\bibinfo{volume}{23715}}
  (\bibinfo{year}{2019}{\natexlab{c}}).

\bibitem[{\citenamefont{{Mirzoyan} et~al.}(2019)\citenamefont{{Mirzoyan},
  {Noda}, {Moretti}, {Berti}, {Nigro}, {Hoang}, {Micanovic}, {Takahashi},
  {Chai}, {Moralejo} et~al.}}]{2019GCN.23701....1M}
\bibinfo{author}{\bibfnamefont{R.}~\bibnamefont{{Mirzoyan}}},
  \bibinfo{author}{\bibfnamefont{K.}~\bibnamefont{{Noda}}},
  \bibinfo{author}{\bibfnamefont{E.}~\bibnamefont{{Moretti}}},
  \bibinfo{author}{\bibfnamefont{A.}~\bibnamefont{{Berti}}},
  \bibinfo{author}{\bibfnamefont{C.}~\bibnamefont{{Nigro}}},
  \bibinfo{author}{\bibfnamefont{J.}~\bibnamefont{{Hoang}}},
  \bibinfo{author}{\bibfnamefont{S.}~\bibnamefont{{Micanovic}}},
  \bibinfo{author}{\bibfnamefont{M.}~\bibnamefont{{Takahashi}}},
  \bibinfo{author}{\bibfnamefont{Y.}~\bibnamefont{{Chai}}},
  \bibinfo{author}{\bibfnamefont{A.}~\bibnamefont{{Moralejo}}},
  \bibnamefont{et~al.}, \bibinfo{journal}{GRB Coordinates Network, Circular
  Service, No.~23701, \#1 (2019)} \textbf{\bibinfo{volume}{23701}}
  (\bibinfo{year}{2019}).

\end{thebibliography}

\end{document}